\documentclass[graybox]{svmult}

\usepackage{mathptmx}       
\usepackage{helvet}         
\usepackage{courier}        
\usepackage{type1cm}        
\usepackage{makeidx}         
\usepackage{graphicx}        
\usepackage{multicol}        
\usepackage[bottom]{footmisc}

\makeindex             

\usepackage{smartdiagram}
\usepackage{bbm}
\usepackage{amsmath,amssymb,xcolor,pxfonts}
\usepackage{tabularx,booktabs}
\newtheorem{goal}{Problem}
\usepackage{tikz}
\usepackage{listofitems} 
\usetikzlibrary{arrows.meta} 
\usepackage[outline]{contour} 
\contourlength{1.4pt}

\tikzset{>=latex} 
\usepackage{xcolor}
\colorlet{myred}{red!80!black}
\colorlet{myblue}{blue!80!black}
\colorlet{mygreen}{green!60!black}
\colorlet{myorange}{orange!70!red!60!black}
\colorlet{mydarkred}{red!30!black}
\colorlet{mydarkblue}{blue!40!black}
\colorlet{mydarkgreen}{green!30!black}
\tikzstyle{node}=[thick,circle,draw=myblue,minimum size=22,inner sep=0.5,outer sep=0.6]
\tikzstyle{node in}=[node,green!20!black,draw=mygreen!30!black,fill=mygreen!25]
\tikzstyle{node hidden}=[node,blue!20!black,draw=myblue!30!black,fill=myblue!20]
\tikzstyle{node convol}=[node,orange!20!black,draw=myorange!30!black,fill=myorange!20]
\tikzstyle{node out}=[node,red!20!black,draw=myred!30!black,fill=myred!20]
\tikzstyle{connect}=[thick,mydarkblue] 
\tikzstyle{connect arrow}=[-{Latex[length=4,width=3.5]},thick,mydarkblue,shorten <=0.5,shorten >=1]
\tikzset{ 
  node 1/.style={node in},
  node 2/.style={node hidden},
  node 3/.style={node out},}
\def\nstyle{int(\lay<\Nnodlen?min(2,\lay):3)} 

\begin{document}

\title*{The outcomes of  generative AI are exactly the Nash equilibria of a non-potential game}
\titlerunning{ Generative AI Outcomes are non-Potential Game  Equilibria}
\author{Boualem Djehiche and Hamidou Tembine}
\institute{B. Djehiche (corresponding author) \at Department of Mathematics,  KTH Royal Institute of Technology,\\ 10044, Stockholm, Sweden. \\ ORCID number: 0000-0002-8171-9599\\ \email{boualem@kth.se}\and
H. Tembine \at 
Learning and Game Theory Laboratory, TIMADIE, Paris 75116, France\\ ORCID number: 0000-0002-1604-8223\\
\email{tembine@landglab.com}}
%

\maketitle

\abstract{In this article we show that the asymptotic outcomes of both shallow and deep neural networks such as those used in BloombergGPT to generate economic time series are exactly the  Nash equilibria of a non-potential game.  We then design and analyze deep neural network algorithms that converge to these equilibria. The methodology is extended to federated deep neural networks between clusters of regional servers and on-device clients. Finally, the variational inequalities behind large language models including  encoder-decoder related transformers are established.}

\section{Introduction}
Designing   neural networks that  are stable and accurate  is a challenging task \cite{reflimit}. Due to the composition of multiple activation operators and weight operators, neural networks generate non-convex functions  with respect to the weights and sometimes with respect to the initial inputs. As a consequence, vanishing sub-differential (Fermat's rule) may include local extrema, saddle points and other unstable neighborhoods. By looking at the training problem and neural network asymptotic outcomes not from a minimization perspective but from a variational inequalities or from a fixed points perspective, one can obtain a plausible explanation of the possible  asymptotic outcomes. To this end, what is inside the black-box needs to be examined in details. 

A strong motivation of the present work is to better understand the architecture BloombergGPT  uses in economic time series and studied in \cite{BloombergGPT} as a major example of outcomes of generative AI.

\subsection{Related work}
The connection between deep learning and game theory has been examined in several recent papers. The work \cite{nopot2} surveys the interplay between the two fields. On the one hand, Game theory has developed different solution concepts. Many of them involve fixed-point arguments and variational inequalities. In a game theoretic setting, the objective functions do not have to be convex (in the global variable) and there are several strategic learning algorithms  for solving game-theoretic problems beyond the class of potential convex games. On the other hand, deep neural networks input-output outcomes \cite{th1,th2,th3} are fixed-points of a composition of activation, attention, and weight operators. 
The composition  of multiple operators may not be convex in the weight parameters which need to be designed during the training phase of the neural network. Several {\it convex optimization tools} have been used to address the training problem which is a {\it  non-convex optimization} problem in the weight parameters. As a consequence, most of the error bounds, stability  and accuracy analysis provided in the literature  are not exploitable for the training problem.  The  convexity in the input signal variable does not necessarily mean or imply convexity in the weight parameters which also differs from the convexity  of the functionals involved in another problem at hand.


In order to capture the behavior of neural networks, we shall use the features of what is inside the black-boxes layer per layer, data per data, client device per client device, cluster per cluster and server per server by means of a fixed-point argument and variational inequalities associated to each problem. 

{\it  The works in \cite{cc1,cc2} have established properties of activation functions used in neural networks and used  averaged-operators 
to prove convergence in the weak sense. We will build the  large learning  model (LLM) properties based on these references to establish connections with  Stackelberg solution and Nash equilibria. }

\subsection{Contribution of this work} 

In this paper we examine the asymptotic behavior of both deep and shallow neural network architectures (see e.g. \cite{nopot2} for a precise definition) by means of game-theoretic designs and incentives.  We start with the first building blocks which are the activation operators involved in each layer of the neural network. The activation operators are not necessarily convex. Some of them, e.g. ReLU,  are convex functions but still the composition is non-convex in the weight parameters. For $40$ of the most used activation functions we provide a common property that is $\gamma$-averagedness (see Definition \ref{lambda-av} below) for a certain $\gamma\in (0,1]$.  Moreover, for $\gamma$-averaged operators that have at least one fixed-point, there is a class of learning algorithms that converge to that set of fixed-points.  We propose  game-theoretic algorithms that connect to the fixed-points of the deep neural networks.
Our contribution can be summarized as follows.
Our first goal is to look at what is inside the designed boxes of deep neural network architectures from an asymptotic analysis perspective. We show that  the outcomes of neural networks are exactly the  Nash equilibria of a game. Moreover, the corresponding class of games cannot be reduced to the class of potential games. The involved functionals are not necessarily convex w.r.t.  the weights and  the inputs. Given the structure of neural networks, a multi-layer  Stackelberg game emerges as a solution concept. We connect the Nash equilibria and the Stackelberg solution of the corresponding game. We provide strategic learning algorithms for the corresponding solution concepts associated to the asymptotic network outcomes and for the training problem. These algorithms are robust to (input, bias, weight, activation and attention) perturbations as well as to measurement noise and  numerical errors under suitable summability conditions. Our second goal is to look at what is inside the designed boxes of strategic deep learning from an equilibrium analysis perspective. We examine  federated deep learning with clustered servers and several clients per server, and characterize the fixed-points of federated learning in terms of training outcomes as well as in terms of architecture outcomes.

\subsection{Relevance to economic time series}

The architecture of a large  learning model for economic time series incorporates key mathematical elements such as  neural networks  to capture temporal dependencies, attention mechanisms for focusing on critical time steps and data modalities, graph neural networks to model economic relationships, autoencoders for dimensionality reduction, fitting, filtering, and classification techniques for forecasting and trend prediction, regularization methods to prevent overfitting, strategic learning algorithms  for efficient training, customized loss functions for each economic agent to reflect domain-specific knowledge, ensemble learning to enhance prediction performance, and interpretability techniques for insight into the model's decision-making process, enabling accurate forecasting and insightful analysis of complex economic dynamics. As an example, BloombergGPT \cite{BloombergGPT}  is specialized to economic time series. It is a 50 billion parameter language model that is trained on a wide
range of economic data. They constructed a 363 billion token dataset based on Bloomberg's extensive data sources, perhaps the largest domain-specific dataset yet, augmented with 345 billion tokens from general purpose datasets. Their architecture properties are covered in this paper.

\subsection{Structure}
The rest of the paper is organized as follows. The next section presents the neural network asymptotic outcome problem, its solutions (if any) and the associated game and equilibrium algorithm to find them.  It is then followed by the training problem, its solution (if any) as a Nash equilibrium of a well-designed game and iterative algorithms to find them. 
The last section  extends to federated (deep/shallow) neural networks between regional cluster servers and clients.

\section{Formulation of the problem}
\begin{figure}[htb]
\centering
\begin{tikzpicture}[x=1.5cm,y=1cm]
  \message{Neural network, shifted}
  \readlist\Nnod{4,7,8,6,4} 
  \readlist\Nstr{d,m,n,p,d} 
  \readlist\Cstr{\strut x,O^{(\prev)},O^{(\prev)},O^{(\prev)},y} 
  \def\yshift{0.5} 
  
  \message{^^J  Layer}
  \foreachitem \N \in \Nnod{ 
    \def\lay{\Ncnt} 
    \pgfmathsetmacro\prev{int(\Ncnt-1)} 
    \message{\lay,}
    \foreach \i [evaluate={\c=int(\i==\N); \y=\N/2-\i-\c*\yshift;
                 \index=(\i<\N?int(\i):"\Nstr[\lay]");
                 \x=\lay; \n=\nstyle;}] in {1,...,\N}{ 
      \node[node \n] (N\lay-\i) at (\x,\y) {$\Cstr[\lay]_{\index}$};
      
      \ifnum\lay>1 
        \foreach \j in {1,...,\Nnod[\prev]}{ 
          \draw[connect,white,line width=1.2] (N\prev-\j) -- (N\lay-\i);
          \draw[connect] (N\prev-\j) -- (N\lay-\i);
        }
      \fi 
      
    }
    \path (N\lay-\N) --++ (0,1+\yshift) node[midway,scale=1.5] {$\vdots$};
  }
  
  \node[above=5,align=center,mygreen!60!black] at (N1-1.90) {input\\[-0.2em]layer};
  \node[above=2,align=center,myblue!60!black] at (N3-1.90) {hidden layers};
  \node[above=10,align=center,myred!60!black] at (N\Nnodlen-1.90) {output\\[-0.2em]layer};
  
\end{tikzpicture}\\
\caption{A schematic representation of the deep learning architecture considered in this paper. $O^{(l)}_k:=O_{l,t}(k)$ displayed in Eq. \eqref{O-l-t}. The integers $m,n,p,d $ may be different.}
\label{fig:anny}
\end{figure}
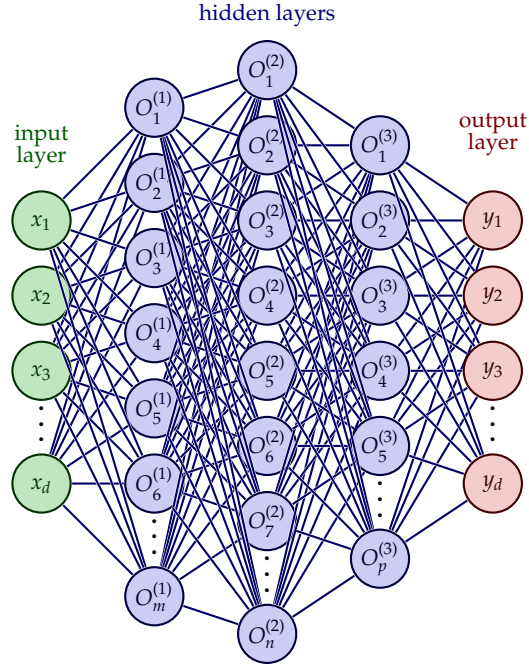

We consider a  neutral network with $L\geq 1$ layers (see Figure \ref{fig:anny}). Let $\{\mathcal{H}_l\}_{0\leq l\leq L}$ be non-zero Hilbert spaces. For every time $t\in \mathbb{N}$, and every $l\in \{1,\ldots,L\},$ let $W_{l,t}: \ \mathcal{H}_{l-1} \rightarrow \mathcal{H}_l$ be a bounded linear operator and consider the family of activation operators $R_{l,t}: \ \mathcal{H}_{l} \rightarrow \mathcal{H}_l$. The deep neural network is defined by the composition of operators
\begin{equation}\label{comp-op}
R_{L,t}\circ (W_{L,t}\cdot+b_{L,t})\circ R_{L-1,t}\circ (W_{L-1,t}\cdot+b_{L-1,t})\circ \ldots  \circ R_{1,t}(W_{1,t}\cdot+b_{1,t}),
\end{equation}
where $W_{l,t}$ is the weight operator and $b_{l,t}$ captures the bias parameter. Let  $x_0\in \mathcal{H}_0,$
$\{\lambda_t\}_{t\geq 0}$ be a non-negative sequence and  consider the map define by
\begin{align}\label{O-l-t}
    O_{l,t}: \ & \mathcal{H}_{l-1} \rightarrow \mathcal{H}_l  \nonumber \\
    & x \mapsto  R_{l,t}(W_{l,t}x+b_{l,t}).
\end{align}
Iterate layer by layer and by timestep the following:
\begin{equation} \label{dl1}
     \begin{array}{l|l}
    \mbox{for}\ t\in \{ 0,1, 2,\ldots\} & 
    y_{1,t}=O_{1,t}(x_t),\\  &
    \mbox{for}\ l\in \{2, \ldots, L\},\ 
    y_{l,t}=O_{l,t}(y_{l-1,t}),\\ &  x_{t+1}= x_{t}+\lambda_t (y_{L,t}-x_t).
\end{array} 
\end{equation}
which means that  $$x_{t+1}=x_t+\lambda_t(O_{L,t}\circ O_{L-1,t}\circ \ldots \circ O_{1,t}(x_t)-x_t), $$ starting from $x_0.$ For consistency of the operation, the output of $O_{L,t}$ should be in the same space as the input i.e., $\mathcal{H}_L=\mathcal{H}_0.$

\begin{definition} The neural network specification is given by the family $$
\mathbb{NN}:=(x_0,R,W,b,L, \{\mathcal{H}_l\}_{1\leq l\leq L},\lambda).
$$
\end{definition}

In this paper we address the following two problems for all the 40 activation functions $R:=(r_k)_k$ currently used in deep neural networks architectures and which are displayed in Tables \ref{actionaverage1} and \ref{actionaverage2} below.

\begin{table}[htb]
 \caption{Activation functions that are $\gamma$-averaged, $\gamma\in (0,1].$}
\begin{tabular}{| p{4cm} |p{4.2cm} | p{2.5cm}|} \toprule
Activation function & Expression &  $\gamma$-averagedness \\ \midrule
Identity & $r_1(x)= Id(x)=x$ & 1 \\  \hline
Linear &  $r_2(x)= \langle \lambda, x \rangle +b$ & $ \frac{(1+\|\lambda\|_{\infty})}{2} \mathbb{I}_{\{\|\lambda\|_{\infty} \leq  1\}} $\\ \hline
Rectified linear unit (ReLU) & $r_3(x)=\max(0,\langle \lambda, x \rangle +b) $ & $ \frac{(1+\|\lambda\|)}{2} \mathbb{I}_{\{\|\lambda\|\leq  1\}} $ \\ \hline
 Logistic (Sigmoid, Soft step)  & $r_4(x)=\frac{1}{1+exp(-\langle \lambda, x \rangle -b)} $& $\frac{(4+\|\lambda\|)}{8} \mathbb{I}_{\{\|\lambda\|\leq  1\}} $ \\  \hline
  Sigmoid  & $\sigma(x)=sigmoid(x)=\frac{1}{1+e^{-x }} $& $\frac{5}{8}$ \\  \hline
  Hyperbolic tangent &  $r_5(x)= \lambda\frac{e^{x}-e^{-x}}{e^{x}+e^{-x}}$ &  $ \frac{(1+\|\lambda\|)}{2} \mathbb{I}_{\{\|\lambda\|< 1\}} $  \\  \hline
  Softmax &$r_6(x)= \frac{ e^{\lambda x_i}}{\sum_{k=1}^K e^{\lambda x_k}}$& $\frac{(1+\|\lambda\|)}{2} \mathbb{I}_{\{\|\lambda\|< 1\}} $ \\ \hline
  Gaussian error linear unit (GELU2) & $\begin{array}{lll}
 r_7(x)=\lambda \frac{1}{2} x\left[ 1+tanh\left(\sqrt{\frac{2}{\pi}}(x \right.\right. \\ \left.\left. \qquad + 0.044715 x^3)\right) \right]\end{array}$ & $\frac{18}{20}$ \\ \hline
   Gaussian error linear unit & $GELU(x)=\lambda(x) \mathbb{P}(X\leq x)= \lambda(x) \frac{1}{2}\left[ 1+erf(\frac{x}{\sqrt{2}})\right],\  X\sim \mathcal{N}(0,1)$ &  $ \frac{(1+\|\lambda\|)}{2} \mathbb{I}_{\{\|\lambda\|\leq  1\}} $  \\ \hline
  Softplus & $r_8(x)=\lambda \log(1+e^{x})$&  $ \frac{(1+\|\lambda\|)}{2} \mathbb{I}_{\{\|\lambda\|\leq  1\}} $  \\ \hline
  Softplus & $softplus(x)= \frac{1}{\lambda}\log(1+e^{\lambda x})$&1 \\ \hline
  & $softplus(x)= \log(1+\sum_{k=1}^d e^{x_k})$& 1\\ \hline
  Exponential linear unit (ELU) & $r_9(x)=\lambda (e^{x}-1) \mathbb{I}_{\{x\leq 0\}}+ x\mathbb{I}_{\{x> 0\}}$& 1 \\ \hline
  Scaled exponential linear unit &$r_{10}(x)=
  \lambda [\alpha (e^{x}-1) \mathbb{I}_{\{x< 0\}}+ \alpha x\mathbb{I}_{\{x\geq  0\}}],\ \ (\alpha,\lambda)=(0.0507,0.6733)$& $\lambda \alpha$\\ \hline
  Leaky rectified linear unit &$r_{11}(x)= 0.01 x \mathbb{I}_{\{x< 0\}}+ x\mathbb{I}_{\{x\geq 0\}}$& 1\\ \hline
  Parametric rectified linear unit &$r_{12}(x)=\lambda x \mathbb{I}_{\{x< 0\}}+ x\mathbb{I}_{\{x\geq 0\}}$& 1 \\ \hline
  Sigmoid linear unit (Sigmoid shrinkage) &$r_{13}(x)=\frac{x}{1+e^{-x}}$& 1\\ \hline
  Swish  &$r_{14}(x)= \epsilon x\  sigmoid(\lambda x)$ &
 $\frac{10+11\epsilon}{20}$ \\ \hline
  Gaussian &$r_{15}(x)= e^{-\langle x,x\rangle}$&  $\frac{1+e^{-1}}{2} $\\ \hline
  Maxout &$r_{16}(x)= \max_{k}x_k$& 1 \\ \hline
  Approximate Heaviside / Binary step  &$r_{17}(x)= \sigma(x/\epsilon)$ & $\frac{(1+4\epsilon )}{8\epsilon} \mathbb{I}_{\{\epsilon\geq  1/4\}} $\\ \hline
  Multiquadratics & $r_{18}(x)= \sqrt{ (x-\alpha)^2+\lambda^2}$ & 1\\ \hline
  Inverse multiquadratics &$r_{19}(x)=\frac{1}{\sqrt{ (x-\alpha)^2+(1+\lambda)^2}}$& $\frac{2+\lambda}{2(1+\lambda)}$ \\ \hline
  Mish & $r_{20}(x)= x \ tanh( \ softplus(x))$ & see Lemma \ref{lemmann}  \\ \hline
   \bottomrule
  \end{tabular}
  \label{actionaverage1}
\end{table}

\begin{table}[htb]
\caption{Activation functions   that are $\gamma$-averaged, $\gamma\in (0,1]$ (cont.)}
\begin{tabular}{| p{4cm} |p{4.2cm} | p{2.5cm}|} \toprule
  Metallic mean  & $r_{21}(x)=\frac{x+\sqrt{x^2+4}}{2}$& $\frac{1}{2}$ \\ \hline
  Arc tangent &  $r_{22}(x)= tan^{-1}(x)$ & 1 \\ \hline
Softsign&  $r_{23}(x)= \frac{x}{1+|x|}$ & 1 \\ \hline
Inverse square root unit&  $r_{24}(x)= \frac{x}{\sqrt{1+(1+\lambda) x^2}}$ & $\frac{1+\sqrt{1+\lambda }}{2\sqrt{1+\lambda}}$\\ \hline
Inverse square root linear unit&  $r_{25}(x)= \frac{x}{\sqrt{1+\lambda x^2}}\ \mathbb{I}_{\{x<0\}}+ x \mathbb{I}_{\{x\geq 0\}}$ & 1 \\ \hline
Square nonlinearity &  $r_{25}(x)=  -\mathbb{I}_{\{x< -2\}}+
 (x+\frac{x^2}{4})\mathbb{I}_{\{-2\leq x< 0\}}+ (x-\frac{x^2}{4})\mathbb{I}_{\{0\leq x\leq 2\}}+ \mathbb{I}_{\{x> 2\}}$ &1 \\ \hline
Bent identity &  $r_{26}(x)=\frac{2}{3}\lambda( x+\frac{-1+\sqrt{1+x^2}}{2})$ & $\lambda$ \\ \hline
Softexponential&  $r_{27}(x)=  -\frac{\log(1-\lambda(x+\lambda))}{\lambda}\mathbb{I}_{\{\lambda< 0\}}+
 x\mathbb{I}_{\{\lambda=0\}}+ (\lambda+\frac{e^{\lambda x}-1}{\lambda})\mathbb{I}_{\{\lambda>0\}}$ & 1\\ \hline
 Soft clipping &  $r_{28}(x)= \frac{1}{\lambda}\log( \frac{1+e^{\lambda x}}{1+e^{\lambda(x-1)}})$ & \\ \hline
  &  $\tilde{r}_{28}(x)= ( \frac{x_k}{1+\|x\|})_k$ & see $r_{23}$\\ \hline
Sinusoid &  $r_{29}(x)= sin(x)$ & 1 \\ \hline
Sinc &  $r_{29}(x)= \frac{sin(x)}{x}\mathbb{I}_{\{x\neq  0\}} + \mathbb{I}_{\{x=  0\}}$ & 1\\ \hline

Piecewise linear &  $r_{30}(x)=  0 \mathbb{I}_{\{x\leq  -\frac{1}{2}\}}+
 (x+\frac{1}{2})\mathbb{I}_{\{-\frac{1}{2}< x< \frac{1}{2}\}}+\mathbb{I}_{\{ x>\frac{1}{2} \}}$ & 1\\ \hline
    Sinu-sigmoidal Linear Unit  &$r_{32}(x)=(x+\lambda \sin(\alpha x))\ sigmoid(x)$ & see Lemma \ref{lemmann}  \\ \hline
   Complementary Log-Log&$r_{33}(x)= 1-e^{-e^{x}}$ & 3/4 \\ \hline
 Bipolar Sigmoid&$r_{34}(x)= \frac{1-e^{-x}}{1+e^{-x}}$ & see $r_5$\\
\hline

Hard Tanh&$r_{35}(x)= \max(-1,min(1,x))$ &  1\\ \hline
 Absolute value &$r_{36}(x)= |x|$ &  1 \\ \hline
  Logit&$r_{36}(x)=\frac{1}{10}\log(\frac{x}{1-x})\mathbb{I}_{[1/4, 3/4]}(x)$ & 3/4\\ \hline
   Softsign( Probit)&$r_{37}(x)= softsign(\Phi^{-1}(x)),\ \Phi(x)= \frac{1}{2}\left[ 1+erf(\frac{x}{\sqrt{2}})\right]$ & $\frac{9}{10}$ \\ \hline
    Linear Gaussian&$r_{38}(x)= x e^{-x^2}$ & $ \frac{2+ \sqrt{e}}{4} $ \\ \hline
    Attention-based &$r_{39}=  softmax \circ r_0 $ &  see Lemma \ref{lemmann}  \\ \hline
Attention-based &$r_{40}=r_{38}\circ   softmax \circ r_0 $ & see Lemma \ref{lemmann}  \\ \hline
   \bottomrule
  \end{tabular}
  \label{actionaverage2}
\end{table}

\medskip

\begin{goal}  What does a well-trained Deep Neural Network Do?  To answer this question, we aim to find and characterize the asymptotic behavior of the network for a large class of architectures 
$\mathbb{NN}$.  To this end, we examine the possible limit (if any) of  $\lim\, (x_t, y_{1,t},\ldots, y_{L,t})$  as $t$ goes to infinity.
\end{goal}

\begin{goal}[Training \& Design]  Given an input-output data set $\{(x_t,d_t=y_{L,t}), \ t\in \{1,\ldots, T\}\},$  Find 
$\mathbb{NN}$  such that  $  O_{L,t}\circ O_{L-1,t}\circ \ldots \circ O_{1,t}(x_t)=d_t,\  \forall t,$ 
which is reformulated in a weak sense as an evolving variational inequality: Find a vector $\theta^*:=(W^*,b^*)=(W^*_{l,t}, b^*_{l,t})_{l,t}$  such that, with $\theta_{l,t}:=(W^*_{l,t}, b^*_{l,t})$,
\begin{equation} \label{trainingvarineq}
\sum_{t=1}^T \omega_{l,t}  \Big\langle (R_{l,t}[ A_{l,t} \theta_{l,t}^* ]-y_{l,t}), A_{l,t}( \theta_{l,t} -\theta_{l,t}^*)\Big\rangle \geq 0, \,\,\, \forall \theta, \,\, \forall l\in \mathcal{L},
\end{equation} where $\omega_{l,t}> 0,\ \sum_{t=1}^T \omega_{l,t}=1,\ $  $ \ A_{l,t}: \ ( W_{l,t} , b_{l,t}   ) \mapsto W_{l,t} x_{l,t} +b_{l,t}=A_{l,t}\theta_{l,t},$ is linear  and non-zero, and $A_{l,t}^*$ is the adjoint operator of $A_{l,t} $. Moreover, design an algorithm that approximates a $\theta^*$ , a  solution (if any) to \eqref{trainingvarineq}.
%
%
\end{goal}

\begin{proposition}\label{refbanachann} Consider a deep neural network $(x_0,R,W,b,L, \{\mathcal{H}_l\}_{1\leq l\leq L})$ with $R_{l,t}$  being $1$-Lipschitz continuous (i.e. with Lipschitz constant 1) for all layers (input, hidden and output) and time, for any $k\in \{1,\ldots, L\}$ and $T \geq 0$, where
$$
\underset{t\in \{1,\ldots,T\}}{\max}\,\, \underset{l\in \{1,2,\ldots, k\}}{\max}\, \| W_{l,t}\| < 1.
$$  
Then, independently of the starting signal input $x_0$, there is a strong convergence of the sequence $x_t$ in (\ref{dl1}) to a unique fixed-point for $\lambda_t=1\, \forall t.$ Moreover, the convergence rate is linear.
\end{proposition}

{\bf Proof:} The proof follows from Banach-Picard fixed point theorem with the strict contraction property.

Proposition \ref{refbanachann} imposes a strong condition on the weights. This condition is too restrictive in terms of the norms of weight operators. One can relax the condition using $\gamma$-averaged operators (to be defined below) and hence allowing more flexibility in the design of weights in the deep neural network architectures.

\subsection{Fixed-point of $\gamma$-averaged operators}

\begin{definition}\label{lambda-av} Let $\gamma\in (0, 1].$ An operator $O:\ \mathcal{H}_0 \rightarrow \mathcal{H}_0$  is $\gamma$-averaged if $[Id+\frac{1}{\gamma}(O-Id)]$ is $1$-Lipschitz continuous.
\end{definition}

In the next proposition we state the main properties satisfied by all the 40 activation functions $r_k$ displayed in Tables \ref{actionaverage1} and \ref{actionaverage2}. These properties are easy to check on a case-by-case basis using e.g. Maple or Wolfram Mathematica.

\begin{proposition}\label{annthm2}
For each $k\in \{1,2,\ldots,40\}$,
\begin{itemize}
\item[(i)]\,\, $r_k$ is a $\gamma$-averaged operator. 
\item[(ii)]\,\,   
When $\gamma<1$, the operator $r_k$ can be written as
\begin{equation}\label{r-k-f}
    r_k=[Id+\partial f_k]^{-1}=\arg\min_{y}(f(y)+\frac{1}{2}\|\cdot-y\|^2)
\end{equation}   for some proper lower semi-continuous convex function $f_k$, where $\partial f_k(x):=\{ p, \,\, \ \forall y,\, \langle x-y, p\rangle +f_k(x)\leq f_k(y) \}$ denotes the subdifferential of $f_k$. Moreover, we have
\begin{equation}\label{r-k-f-1} 
r_k=\partial (\frac{1}{2} \| \,\cdot\,\|^2  + f_k)^*,
\end{equation}
where $\psi^*(x)=\sup_{y} [\langle x,y\rangle -\psi(y)] $ is the Legendre-Fenchel conjugate of $\psi$.
 \item [(iii)]\,\, $Fix(r_k)=\arg\min (f_k)$, i.e. the fixed-points of $r_k$ are the minimizers of  $f_k$. \end{itemize}
\end{proposition}
For a proof of the second equality in (ii) see Proposition 6.a. in \cite{Moreau1965} (see also the equality (16.31) in \cite{bc}). By  Corollary 16.24 in \cite{bc}, the equality \eqref{r-k-f-1} follows from  
$$
[ \partial(\frac{1}{2} \| \,\cdot\,\|^2  + f_k)]^{-1}=\partial[(\frac{1}{2} \| \,\cdot\,\|^2  + f_k)^*],
$$
where $\psi^*(x)=\sup_{y} [\langle x,y\rangle -\psi(y)] $ is the Legendre-Fenchel conjugate of $\psi$.  For a proof of (iii) see Proposition 12.28 in \cite{bc}. 

\begin{remark} 
For the activation function  $r_1(x)= Id(x)=x$, $\gamma=1$ and  $f_1(x)=0$.

For the activation function $r_4(x)=\frac{1}{1+e^{-x }}$, $\gamma=\frac{5}{8}$ and 
\begin{equation} \label{f4}
   f_4(x)=   \left\{\begin{array}{ll}
   \frac{1+2x}{2}\log ( \frac{1+2x}{2}) + \frac{1-2x}{2}\log (\frac{1-2x}{2})-\frac{4x^2+1}{8}\  & \mbox{if}\  |x| <\frac{1}{2},\\
   -\frac{1}{4} \  & \mbox{if}\  |x| =\frac{1}{2},\\
   +\infty  & \mbox{if}\  |x| >\frac{1}{2}.\\  
\end{array} \right.
\end{equation}

 For the activation function softsign $r_{23}$ defined by $r_{23}(x)=\frac{x}{1+|x|}$,  $\gamma=1$ and
\begin{equation} \label{f23}
   f_{23}(x)=   \left\{\begin{array}{ll}
 -|x|-\log(1-|x|)-\frac{x^2}{2}  \ &  \mbox{if}\  |x| < 1,\\
   +\infty \ \ \ &  \mbox{if}\  |x| \geq 1.
\end{array} \right.
\end{equation}

\end{remark}

\medskip
Our analysis uses the following results.
\begin{lemma}\label{lemmann00}
Let $(\rho_t,\mu_t,\epsilon_t)$ be nonnegative sequences  such that $\sum_{t\geq 0} \epsilon_t <+\infty$ 
and $\rho_{t+1}\leq \rho_t-\mu_t+\epsilon_t.$ Then,  $\{\rho_t\}_t$ converges and  $\sum_{t\geq 0}\mu_t < \infty$.
\end{lemma}
The proof of Lemma \ref{lemmann00} is straightforward.

\begin{lemma} [see Theorem 3.1 in \cite{nopot} and Chap. 5 in \cite{bc}] \label{lemmann}
 Let $L\geq 1$ be an integer and consider the sequence $0<\gamma_l\leq 1,\  l\in \{1,\ldots,L\}.$ 
\begin{itemize}

\item[(i)]\,\, Let $0<\gamma\leq 1$. Then $[Id+\frac{1}{\gamma}(O-Id)]$ is $1$-Lipschitz continuous if and only if
$$
\forall (x,y)\in \mathcal{H}_0^2,\,\,\,  \| O(x)-O(y) \|^2 \leq \| x-y\|^2-\frac{(1-\gamma)}{\gamma}\| x-O(x)-y+O(y) \|^2.
$$

\item[(ii)]\,\, The composition  $O_L\circ O_{L-1}\circ \ldots O_1$ of $ \gamma_l$-averaged  operators $O_l:\ \mathcal{H}_0 \rightarrow \mathcal{H}_0$ is  $\frac{1}{1+\frac{1}{\sum_{l=1}^L \frac{\gamma_l}{1-\gamma_l}}}$-averaged.
\item[(iii)]\,\, Assume each $O_l$ is a $\gamma_l$-averaged operator. Let $\omega_l\geq 0$ such that $\sum_{l=1}^L \omega_l= 1$. Then the weighted sum operator $\sum_{l=1}^L \omega_l O_l$ is $\sum_{l=1}^L \omega_l\gamma_l$-averaged.
\item[(iv)]\,\, If $O$ is $\mu$-Lipschitz continuous with $\mu<1.$ Then $O$ is $\frac{(1+\mu)}{2}$-averaged.
\item[(v)]\,\, Let $\text{Fixed}(O)=\{x\in \mathcal{H}_0 | \ x=O(x)\}$ be the set of fixed-points of the operator $O.$ Assume that $\cap_{l=1}^L \text{Fixed}(O_l)\neq \emptyset.$  Then 
$$ \text{Fixed}(\sum_{l=1}^L \omega_l O_l)=\text{Fixed}( O_L\circ O_{L-1}\circ \ldots O_1)=\cap_{l=1}^L \text{Fixed}(O_l).$$ 
\item[(vi)]\,\, (Propositions 5.14 and 5.15 in \cite{bc}) Let $0<\gamma\leq 1,\ $ and $O:\ \mathcal{H}_0 \rightarrow \mathcal{H}_0$ be an $\gamma$-averaged operator such that $\text{Fixed}(O)\neq \emptyset.$ Let $\{\gamma_t\}_{t\geq 1}$ be a sequence in $[0,\frac{1}{\gamma}]$ such that $\sum_{t\geq 1}\ \gamma_t(1-\gamma\cdot \gamma_t)=+\infty$. Assume that $x_0\in \mathcal{H}_0$ and set $x_{t+1}=x_t+\gamma_t(O(x_t)-x_t).$ Then $O(x_t)-x_t$ converges  to $0$ as $t$ goes to infinity and $x_t$ converges weakly to a point in $\text{Fixed}(O)$.
\end{itemize}
\end{lemma}

{\bf Proof:}
We start by proving  Lemma \ref{lemmann} (i).
Let $0<\gamma\leq 1. $  The operator $[Id+\frac{1}{\gamma}(O-Id)]$ is $1$-Lipschitz if
$$\| (1-\frac{1}{\gamma}) x+\frac{1}{\gamma} O(x)
- (1-\frac{1}{\gamma}) y-\frac{1}{\gamma} O(y) \| \leq \|x-y\|.$$
Consider the quantity
\begin{equation} \label{dl1t3}
     \begin{array}{llll}
     0\leq \gamma[\|x-y\|^2 -\| (1-\frac{1}{\gamma}) x+\frac{1}{\gamma} O(x)
- (1-\frac{1}{\gamma}) y-\frac{1}{\gamma} O(y) \|^2]\\
\quad= \gamma[\|x-y\|^2 - \frac{1}{\gamma^2}\| -(1-\gamma) (x-y)+O(x)-O(y) \|^2]\\
=\quad \gamma [\|x-y\|^2 - \frac{1}{\gamma^2}\left((1-\gamma)^2 \|x-y\|^2+\|O(x)-O(y) \|^2 \right. \\ \left. \qquad \qquad -2(1-\gamma)\langle x-y,  O(x)-O(y)\rangle\right)].
     \end{array}
\end{equation}
Using the identity $-2\langle a,b \rangle =\|a-b\|^2-\|a\|^2-\|b\|^2$, we obtain
$$
-2\langle x-y,  O(x)-O(y)\rangle =\| x-O(x)-(y-O(y))\|^2 -\| x-y\|^2-\| O(x)-O(y)\|^2.
$$
 It follows that 
\begin{equation} \label{dl1t4}
    \begin{array}{lll}
     0\leq \gamma[\|x-y\|^2 -\| (1-\frac{1}{\gamma}) x+\frac{1}{\gamma} O(x)
- (1-\frac{1}{\gamma}) y-\frac{1}{\gamma} O(y) \|^2]\\
=\gamma [\|x-y\|^2 - \frac{1}{\gamma^2}\left(  (1-\gamma)^2 \|x-y\|^2+\|O(x)-O(y) \|^2 
\right.  \\ \left. +(1-\gamma)\| x-O(x)-(y-O(y))\|^2 - (1-\gamma)\| x-y\|^2- (1-\gamma)\| O(x)-O(y)\|^2
\right)]
\\ = \| x-y\|^2 - \| O(x)-O(y)\|^2 -\frac{(1-\gamma)}{\gamma}\| x-O(x)-(y-O(y))\|^2.
    \end{array} 
\end{equation}
which completes the proof the $\gamma$-averaged operator inequality.

Lemma \ref{lemmann} (ii) is easily proved by induction. Start with the composition of two operators and apply the recursion.

We now prove (iii). We have

\begin{equation} \label{dl1t5} \begin{array}{ll}
 M_{\gamma}:=Id+\frac{1}{\sum_{k=1}^L \omega_k \gamma_k} \left[\sum_{l=1}^L \omega_l O_l-Id\right]=\sum_{l=1}^L  p_l[ Id+\frac{1}{\gamma_l}(O_l-Id)]
 \end{array}
\end{equation}
where $p_l:= \frac{\omega_l  \gamma_l}{\sum_{k=1}^L \omega_k \gamma_k}.$ Then $p_l> 0$ and $\sum_{l=1}^L p_l=1$. Hence,
$M_{\gamma}$ is a $1$-Lipschitz operator since it is a convex combination of $1$-Lipschitz operators. This proves (iii). 

We now focus on the statement (iv).  We want to show that the operator $M_{\mu}=
Id+\frac{2}{1+\mu}(O-Id)$ $1$-Lipschitz. We have
$M_{\mu}
=\frac{(1-\mu)}{1+\mu} (-Id)+\frac{2\mu}{1+\mu}(\frac{O}{2})$ which is $1$-Lipschitz as a convex combination of  $-Id$ and $\frac{O}{2}$ which are $1$-Lipschitz operators.

The proof of  (v) is elementary. We now focus on the statement (vi). From the assumptions we note that $\epsilon_t=\gamma\cdot\gamma_t$ is a sequence in $[0,1]$ and satisfies  $\sum_{t\geq 1}\epsilon_t (1-\epsilon_t) =+\infty$. Set 
$\widetilde{O}:=(1-\frac{1}{\gamma})Id+\frac{1}{\gamma}O$. Then $x_{t+1}=x_t+\epsilon_t(\widetilde{O}(x_t)-x_t)$ and
$\text{Fixed}(\widetilde{O})=\text{Fixed}(O)$. Let $x^*\in \text{Fixed}(O)$ i.e. $\widetilde{O}(x^*)=O(x^*)=x^*.$ 
We evaluate $\| x_{t+1}-x^*\|$. Indeed,
\begin{equation} \label{dl1t6}
   \begin{array}{lll}
\| x_{t+1}-x^*\|^2
= \|(1-\epsilon_t)x_t+\epsilon_t\widetilde{O}(x_t) - \epsilon_t x^* -(1-\epsilon_t)x^*\|^2\\
=\| (1-\epsilon_t)(x_t-x^*)+\epsilon_t(\widetilde{O}(x_t)-O(x^*))\|^2\\
= (1-\epsilon_t)^2 \| x_t-x^*\|^2+\epsilon^2_t\| \widetilde{O}(x_t)-\widetilde{O}(x^*)\|^2 + 2\epsilon_t(1-\epsilon_t) \langle x_t-x^*,\widetilde{O}(x_t)-\widetilde{O}(x^*)\rangle \\
=(1-\epsilon_t)^2 \| x_t-x^*\|^2+\epsilon^2_t\| \widetilde{O}(x_t)-\widetilde{O}(x^*)\|^2 \\ - \epsilon_t(1-\epsilon_t)  [ \| x_t-\widetilde{O}(x_t)\|^2- \| x_t-x^*\|^2 -\| \widetilde{O}(x_t)-\widetilde{O}(x^*)\|^2]
\\
=[(1-\epsilon_t)^2 +\epsilon_t(1-\epsilon_t) ]\| x_t-x^*\|^2\\
+[\epsilon^2_t +\epsilon_t(1-\epsilon_t) ]\| \widetilde{O}(x_t)-\widetilde{O}(x^*)\|^2
- \epsilon_t(1-\epsilon_t)  \| x_t-\widetilde{O}(x_t)\|^2 \\
=(1-\epsilon_t) ]\| x_t-x^*\|^2\ +\epsilon_t \| \widetilde{O}(x_t)-\widetilde{O}(x^*)\|^2\\  - \epsilon_t(1-\epsilon_t)  \| x_t-\widetilde{O}(x_t)\|^2.
\end{array}
\end{equation}
Therefore, 
\begin{equation} \label{dl1t7}
  \begin{array}{lll}
\| x_{t+1}-x^*\|^2\\ \qquad
= (1-\epsilon_t) ]\| x_t-x^*\|^2\ +\epsilon_t \| \widetilde{O}(x_t)-\widetilde{O}(x^*)\|^2 - \epsilon_t(1-\epsilon_t)  \| x_t-\widetilde{O}(x_t)\|^2\\ \qquad
\leq
\| x_t-x^*\|^2 - \epsilon_t(1-\epsilon_t)  \| x_t-\widetilde{O}(x_t)\|^2.
\end{array} 
\end{equation}
By a telescopic sum, we obtain
\begin{equation} \label{dl1t8}
     \begin{array}{lll}
\sum_{t=0}^T \epsilon_t(1-\epsilon_t)  \| x_t-\widetilde{O}(x_t)\|^2
& \leq 
\sum_{t=0}^T (\|x_{t}-x^*\|^2-\| x_{t+1}-x^*\|^2 ) \\
 & =\|x_{0}-x^*\|^2 -\| x_{T+1}-x^*\|^2.
\end{array} 
\end{equation}

As  $\sum_{t=0}^{\infty} \epsilon_t(1-\epsilon_t)  \| x_t-\widetilde{O}(x_t)\|^2 \leq  \|x_{0}-x^*\|^2 <+\infty$ and  $\sum_{t=0}^{\infty} \epsilon_t(1-\epsilon_t) =+\infty$ it follows  from Lemma \ref{lemmann00} that the sequence \ $\| x_t-\widetilde{O}(x_t)\|^2  \rightarrow 0$ as $t\rightarrow \infty$. But, $O-Id=\gamma(\widetilde{O}-Id)$. Thus, $\| x_t-O(x_t)\|^2  \rightarrow 0$ as $t\rightarrow \infty$.

For a proof of the last statement that the sequence $x_t$ converges weakly to a point in $\text{Fixed}(O)$, we refer to the proof of Theorems 5.15 (iii) in \cite{bc} since it involves further preliminaries that we have chosen to not recall here. This completes the proof.

\begin{remark} $ $ 
\begin{itemize}
\item Lemma \ref{lemmann} (v) assumes the existence of a fixed-point because a $1$-Lipschitz operator does not necessarily admit a fixed-point. It may need invariance and boundedness of the domain so that Brouwer-type fixed-point theorems can be applied. For instance the mapping $x\mapsto x+ 2023$ has no fixed-point in $\mathbb{R}$  but it is a $1$-Lipschitz function. The domain in this particular case is unbounded.
\item For a $\gamma$-averaged operator $O$, the algorithm is given by
\begin{equation} \label{dly1}
     \left\{\begin{array}{l}
     x_0\in \mathcal{H}_0,\\
    \mbox{for}\ t\in \{ 1, 2,\ldots\} \\
x_{t+1}=x_t+\mu_t(O(x_t)-x_t),\\
\end{array} \right.
\end{equation}  where $O$ is not necessarily a contraction and we have more flexibility for the choice of $\mu_t$ up to $\frac{1}{\gamma}\geq 1.$
\end{itemize}
\end{remark}

\begin{remark} $ $
\begin{itemize}
\item The activation operators $r_k$ from Tables \ref{actionaverage1} and \ref{actionaverage2} are not all convex. For example,  take  the sigmoid, the S-shaped function $r_4$ or simply
 $\sigma(x)=\frac{1}{1+e^{-x}}-\frac{1}{2}$. Then 
 $\sigma^{\prime}(x)=\frac{e^{-x}}{(1+e^{-x})^2}$ and 
  $\sigma^{\prime\prime}(x)=\frac{-e^{-x} (1+e^{-x})^2 -e^{-x} 2(-e^{-x})(1+e^{-x})}{(1+e^{-x})^4}
 =\frac{1- e^x  }{ e^{2x} (1+e^{-x})^2}$  The second derivative is $\sigma^{\prime\prime}(x) >0$ for $x<0$,  $\sigma^{\prime\prime}(x)=0$ for $x=0$ and $\sigma^{\prime\prime}(x)<0$ for $x>0.$ This means that on the positive axis branch, the function is not convex: we shift by $\frac{1}{2}$ to obtain
   $\sigma(\log 2)=\frac{2}{3}$ , $\sigma(\log 8)=\frac{8}{9}$
  $$\begin{array}{ll}
  \frac{4}{5}=\sigma(\frac{1}{2}\log 2 + \frac{1}{2}\log 8)\sigma(\log\ 4) > \frac{1}{2} [\frac{2}{3}+\frac{8}{9}]=\frac{14}{18}=\frac{1}{2} [\sigma(\log 2)+\sigma(\log 8)].
  \end{array}
$$
 Thus, $\sigma$ is not convex. However, it is $\frac{5}{8}-$averaged.
\item The activation function softsign $r_{23}$ defined by $r_{23}(x)=\frac{x}{1+|x|}$ is not convex but it is $1$-averaged. 
\item The ReLU activation function is widely used in deep neural networks  and it is given by $r_3(x)=\max(0,x)$  which is convex  and ${1}$-averaged. However the composition $x \mapsto a_1 x+b \mapsto \max(0,a_1x+b_1) \mapsto a_2 \max(0,a_1x+b_1)+b_2 \mapsto \max(0,a_2 \max(0,a_1x+b_1)+b_2)$ is not convex. For training we  consider the mapping $$check_3: ((a_1,b_1), (a_2,b_2) ; (x,y_2)) \mapsto  | y_2-\max(0,a_2 \max(0,a_1x+b_1)+b_2)|^2$$ for the input-output $(x,y_2)$
which reduces to $(a_1,a_2)\mapsto   | 1-\max(0,a_2 \max(0,a_1)|^2$ for $(x,y_2)=(1,1), (b_1,b_2)=(0,0).$
The function $check_3((a_1,0), (a_2,0) ; (1,1))$  evaluated at $(a_1,a_2)=(1,1)$ gives $0$ and at  $(-1,-1)$ gives $1$ meaning that \newline
$check_3((1,0), (1,0) ; (1,1))=0$, 
$check_3((-1,0), (-1,0) ; (1,1))=1$, 
$$\begin{array}{ll}
check_3((0,0), (0,0) ; (1,1))=1 & > \frac{1}{2}(0+1) \\ & = \frac{1}{2} check_3((1,0), (1,0) ; (1,1)) \\ & +
\frac{1}{2} check_3((-1,0), (-1,0) ; (1,1)).
\end{array}
$$
Thus, $check_3$  is not convex in $(a_1,a_2).$ This means that even if ReLU is convex, the output function is not convex in the weight  parameters. Since we optimize the weights when training the neural networks, it leads to non-convex optimization. We will see that there are still some interesting features of the model even if the operator $O_2 \circ O_1$ is non-convex in $(a_1,b_1,a_2,b_2)$ for a given pair $(x,y_2)$.
\end{itemize}
\end{remark}

\begin{remark}  On ReLU activations:
Let $x$ be  input signal $x$ and $y$ be the last layer output.  The following mappings are not convex and not differentiable. 
\begin{itemize}
    \item  $$ (\omega_1,b_1,\omega_2, b_2)\mapsto \mbox{ReLU}(\omega_2 (\mbox{ReLU}(\omega_1*x+b_1))+b_2).$$

\item (x random)  $$ (\omega_1,b_1,\omega_2, b_2)\mapsto \mathbb{E}_{x}\mbox{ReLU}(\omega_2 (\mbox{ReLU}(\omega_1x+b_1))+b_2).
$$ 

\item $$ (\omega_1,b_1,\omega_2, b_2)\mapsto \|\mbox{ReLU}(\omega_2 (\mbox{ReLU}(\omega_1x+b_1))+b_2)-y\|^2.$$ 

\item ($x$ and $y$ random):
$$ (\omega_1,b_1,\omega_2, b_2)\mapsto {\mathbb{E}_{x,y}}\| \mbox{ReLU}(\omega_2 (\mbox{ReLU}(\omega_1x+b_1))+b_2)-y\|^2.
$$ 
\item $$ (\omega,b)\mapsto \frac{1}{D} \sum_{i=1}^D\|\mbox{ReLU}(\omega_2 (\mbox{ReLU}(\omega_1x_i+b_1))+b_2)-y_i\|^2,$$ 
given a set of input signals and layer outputs $\{(x_i,y_i), i\in \{1,\ldots, D\}\}, \ D\geq 1$.
\end{itemize}
\end{remark}

\medskip

We are now ready to state our first main result.
\begin{theorem} [Asymptotic outcome of a neural network] $ $ 
\begin{itemize}
\item The asymptotic outcomes  of the deep neural network $(x_0,R,W,b,L, \{\mathcal{H}_l\}_{1\leq l\leq L})$ are exactly the  Nash equilibria of the non-zero sum game 
$\mathcal{G}=(\mathcal{L}, (\mathcal{H}_l,\  f_l(\cdot)+\frac{1}{2}\| \cdot-b_l-W_lx_{l-1}\|^2)_{l\in \mathcal{L}})$ defined by 
\begin{itemize}
    \item the decision-makers (players) are members of $\mathcal{L}$
    \item the action space of the decision-maker $l$ is $\mathcal{H}_l$
    \item the objective function of decision-maker 
$l$ output is $ z \mapsto f_l(z)+\frac{1}{2}\| z-b_l-W_lx_{l-1}\|^2$

\end{itemize}

\item For three or more layers ($L\geq 3$), there is a deep neural architecture such that the resulting game  is not a potential game.
\end{itemize}
\end{theorem}
Note that this also corresponds to the (multi-level) Stackelberg solution of the game $\mathcal{G}$ as a layer reacts to the previous layer design.

{\bf Proof:}
We start with the first statement of the Theorem.
The set of long-run outcomes of the neural network is the concatenation of all the outputs given by
\begin{equation}  \begin{array}{lll}
x_1^*=O_1 x^*_L,\, x_2^*=O_2\circ O_1 x^*_L,\,\ldots, \,
x_{L-1}^*=O_{L-1}\circ O_{L-2} \ldots O_2\circ O_1 x^*_L, 
\end{array}  
\end{equation}
where $x^*_L \in\mathcal{H}_0$ satisfies $x^*_L=O (x^*_L)$, which is rewritten as 
\begin{equation} \label{outcomedly2}
     \left\{\begin{array}{l}
     x^*=(x^*_1,x^*_2,\ldots,x^*_L) \in \prod_{l=1}^L \mathcal{H}_l,\\
     x_1^*=r_1 (W_1x^*_L+b_1),\\
      x_2^*=r_2 (W_2x^*_1+b_2) ,\\
      \ldots \\ 
       \ldots \\ 
       x_l^*=r_l (W_lx^*_{l-1}+b_l)\\
        \ldots \\ 
       \ldots \\ 
      x_{L-1}^*=r_{L-1}(W_{L-1}x^*_{L-2}+b_{L-1}),\\
       x_{L}^*=r_{L}(W_{L}x^*_{L-1}+b_{L}),\\
\end{array} \right. 
\end{equation}
or, by \eqref{r-k-f},
\begin{equation} \label{outcomedly3}
     \left\{\begin{array}{l}
     x^*=(x^*_1,x^*_2,\ldots,x^*_L) \in \prod_{l=1}^L \mathcal{H}_l,\\
     x_1^*=[Id+\partial f_{1}]^{-1}  (W_1x^*_L+b_1),\\
      x_2^*=[Id+\partial f_{2}]^{-1}  (W_2x^*_1+b_2) ,\\
      \ldots \\ 
       \ldots \\ 
       x_l^*=[Id+\partial f_{l}]^{-1} (W_lx^*_{l-1}+b_l),\\
        \ldots \\ 
       \ldots \\ 
       x_{L}^*=[Id+\partial f_{L}]^{-1} (W_{L}x^*_{L-1}+b_{L}).\\
\end{array} \right. 
\end{equation}



%
Again, by \eqref{r-k-f}, this is equivalent to 
\begin{equation} \label{outcomedly6}
     \left\{\begin{array}{l}
     x^*=(x^*_1,x^*_2,\ldots,x^*_L) \in \prod_{l=1}^L \mathcal{H}_l,\\
   x_1^* \in \arg\min_{y_{1}\in \mathcal{H}_{1}} \ \frac{1}{2}\|y_1-b_1-W_1x^*_L\|^2+f_{1}(y_1) ,\\
        x_2^*\in \arg\min_{y_{2}\in \mathcal{H}_{2}} \    \frac{1}{2}\|y_2-b_2-W_2x^*_1\|^2 + f_{2}(y_2) ,\\
      \ldots \\ 
       \ldots \\ 
       x_l^*\in  \arg\min_{y_{l}\in \mathcal{H}_{l}} \ \frac{1}{2} \| y_l-b_{l}-W_lx^*_{l-1}\|^2+f_{l}(y_l),\\
        \ldots \\ 
       \ldots \\ 
       x_L^*\in \arg\min_{y_{L}\in \mathcal{H}_{0}} \frac{1}{2} \| y_L-b_L-W_{L}x^*_{L-1}\|^2+ f_{L}(y_L).\\
\end{array} \right. 
\end{equation}
By setting  $x^*_0=x^*_L$ we can write
\begin{equation} \label{outcomedly7}
       x_l^*\in  \arg\min_{y_{l}\in \mathcal{H}_{l}} \ \frac{1}{2} \| y_l-b_{l}-W_lx^*_{l-1}\|^2+f_{l}(y_l),\quad
       l\in \{1,\ldots, L\}. 
\end{equation}
In other words
\begin{equation} \label{outcomedly8}
     \begin{array}{l}
   (x_l^*)_{l\in \{1,\ldots, L\}}\ \mbox{is a Nash equilibrium of the non-zero sum game }\ \mathcal{G}.
\end{array} 
\end{equation}
This completes the proof of the first statement. 

We now focus on the second statement with $L\geq 3.$ Since there are at least three layers, due to the cycling behavior, i.e., rotating between these three sets, one can construct a cycle of mappings between these sets as illustrated next. Let $f_l=\mathbb{I}_{S_l}$ where $S_1\neq S_2\neq S_3,$  $S_l$ being not empty, convex and compact and  $\cap_{l=1}^L S_l \neq \emptyset$. Set $W_l=Id$ and $b_l=0$. Then, there is a fixed-point and the set of fixed-points is  $\cap_{l=1}^L S_l.$
However, there is no global best-response potential function associated with the game $\mathcal{G}$ defined above.
{Following \cite{nopot}, there is no global function $P:\  \prod_{l=1}^L \mathcal{H}_l\rightarrow  \prod_{l=1}^L \mathcal{H}_l\ $  whose minimizers are in $\cap_{l=1}^L \mbox{\text{Fixed}}(O_l).$  This means that the game $\mathcal{G}$ is not a best-response potential game}. This completes the proof.

\subsection{Training problem}
We now focus on the training problem  given a data set.  Given $(x_t,y_{L,t})_{t\in \{1,\ldots, T\}}$ the training problem (limited to  weights and biases) 
 is to find  $\theta=(W,b)=(W_{l,t},b_{l,t})_{l,t}$ such that the neural  network $(x_0,r,W,b,L, \{\mathcal{H}_l\}_{1\leq l\leq L})$  produces an output that matches $y_t$ at each iteration time $t.$ This means that  
 $$ 
 (r_L\circ A_{L,\theta_{L,t}})\circ  \ldots \circ (r_1\circ A_{1,\theta_{1,t}})(x_t)=y_{L,t}, \,\,\,  \,t\in \{1,\ldots, T\},
 $$ 
 where $\theta_l:=(W_l,b_l)$ and $A_{l,\theta_{l,t}}=A_{l,(W_{l,t},b_{l,t})}:\ (W_{l,t},b_{l,t}) \mapsto W_{l,t} x_{l-1,t}+b_{l,t}$ with $x_{l-1,t}$ being the output from layer $l-1$.

Thus, given the data $(x_t,y_{L,t})_{t\in \{1,\ldots, T\}}$  the training  problem is  to find 
 $\theta^*=(\theta^*_1,\ldots, \theta^*_L)$ such that 
 \begin{equation} \label{hardform1}
     \begin{array}{l}
   r_L\circ A_{L,\theta^*_{L,t}}\circ  \ldots \circ (r_1\circ A_{1,\theta^*_{1,t}})(x_t)-y_{L,t}=0,\,\,\,  \, t\in \{1,\ldots, T\}.
\end{array} 
\end{equation}

In the sequel we will denote (for each $t$) $A_{l,t}:=A_{l,\theta_{l,t}}$ and  $A_{l,t}^{\dagger}$ be its adjoint operator.
 
\begin{theorem} Suppose that each $r_k$ is $\gamma_k$-averaged for some $\gamma_k\in (0,1), \ 1\leq k\leq L.$ Then,
the  solutions (if any) of the training of weights and biases  problem (\ref{hardform1}) are also solutions of the following variational inequality:
Given an input-output data set $\{(x^*_{0,t}, y^*_{L,t}),\ t\in \{1,2,\ldots, T\}$, find  $ (W^*,b^*)$ such that
\begin{equation}\label{VI-t}
0\in   A^{\dagger}_{L,t}[y_{L,t}^*+\partial f_{L}(y_{L,t}^*)-(W^*_{L,t}x^*_{L-1,t}+b^*_{L,t})], \quad t\in \{1,2,\ldots, T\}.
\end{equation}
These are also Nash equilibria of the non-zero sum game given by
\begin{equation} \label{federatedtraing1x0}
     \left\{\begin{array}{l}
  l\in \mathcal{L},\\
          (W_l^*,b_l^*)\in   \arg\min_{W_l,b_l} \sum_{t=1}^T\omega_{l,t}  \| y_{l,t}^*+\partial f_{l}(y_{l,t}^*)-(W_lx^*_{l-1,t}+b_l)\|^2.
\end{array} \right. 
\end{equation}

\end{theorem}

{\bf Proof:}
Since each of the activation operators $r_k$ satisfies \eqref{r-k-f} (see Proposition \ref{annthm2}), the  training problem is to find $\theta=(W,b)=(W_1,b_1,\ldots,W_L,b_L) $ such that for each $t=1,\ldots,T$,
\begin{equation} \label{federatedtraing1}
     \left\{\begin{array}{l}
  
   (W_{1,t}x^*_{0,t}+b_{1,t}) \in  y_{1,t}^*+\partial f_{1}(y_{1,t}^*),\\
        (W_{2,t}x^*_{1,t}+b_{2,t})\in   y_{2,t}^*+\partial f_{2}(y_{2,t}^*),\\
      \ldots \\ 
       \ldots \\
       (W_{l,t}x^*_{l-1,t}+b_{l,t})\in  y_{l,t}^*+\partial f_{l}(y_{l,t}^*),\\
        \ldots \\ 
       \ldots \\ 
       (W_{L-1,t}x^*_{L-2,t}+b_{L-1,t})\in  y_{L-1,t}^*+\partial f_{L-1}(y_{L-1,t}^*),\\
       (W_{L,t}x^*_{L-1,t}+b_{L,t})\in  y_{L,t}^*+\partial f_{L}(y_{L,t}^*).\\
\end{array} \right. 
\end{equation}
which is rewritten as
\begin{equation} \label{federatedtraing4}
     \left\{\begin{array}{l}
    (W_{1,t}^*, b_{1,t}^*)\in  \arg\min_{W_1,b_1} \| y_{1,t}^* +\partial f_{1}(y_{1,t}^*)-(W_1x^*_{L,t}+b_1) \|^2 ,\\
        (W_{2,t}^*,b_{2,t}^*)\in  \arg\min_{W_2,b_2}  \| y_{2,t}^* +\partial f_{2}(y_{2,t}^*)-(W_2x^*_{1,t}+b_2)\|^2 ,\\
      \ldots \\ 
       \ldots \\ 
          (W_{l,t}^*,b_{l,t}^*)\in   \arg\min_{W_l,b_l}   \| y_{l,t}^*+\partial f_{l}(y_{l,t}^*)-(W_lx^*_{l-1,t}+b_l)\|^2,\\
        \ldots \\ 
       \ldots \\ 
          (W_{L,t}^*,b_{L,t}^*) \in  \arg\min_{W_L,b_L}   \| y_L^*+\partial f_{L}(y_{L,t}^*)-(W_{L}x^*_{L-1,t}+b_{L})\|^2.
\end{array} \right. 
\end{equation}
But, this set is exactly the set of Nash equilibria of the non-zero sum game
$$
\mathcal{G}=\Big(\mathcal{L}, \prod_{l=1}^L L_2( \mathcal{H}_{l-1}, \mathcal{H}_l)\times \mathcal{H}_l, 
 \| y_l^*+\partial f_{l}(y_l^*) -b_l-    W_lx^*_{l-1}\|^2\Big).
 $$
Moreover, by Fermat's principle, the last equation in \eqref{federatedtraing4} implies that
$$
 0\in   A^{\dagger}_{L,t}[y_{L,t}^*+\partial f_{L}(y_{L,t}^*)-(W^*_{L,t}x^*_{L-1,t}+b^*_{L,t})],
 $$
which is the first part of the announced result. 
 
For each  $l\in \mathcal{L}$, let $\omega_{l,t}>0,\ \sum_{t=1}^T \omega_{l,t}=1$. 
Then, clearly the system of variational inequalities \eqref{federatedtraing4} yields
$$
\langle  (W_L,b_L) - (W_L^*,b_L^*), \sum_{t=1}^T  \omega_{L,t} A^{\dagger}_{L,t}[y_{L,t}^*+\partial f_{L}(y_{L,t}^*)-(W^*_{L}x^*_{L-1,t}+b^*_{L})]\rangle \geq 0
$$ for all  $(W_L,b_L)$. Hence, the set of solutions of the training of weights and biases  problem (\ref{hardform1}) is the set of Nash equilibria of the non-zero sum game
\begin{equation} \label{federatedtraing10}
     \left\{\begin{array}{l}
  l\in \mathcal{L},\\
          (W_l^*,b_l^*)\in   \arg\min_{W_l,b_l} \sum_{t=1}^T\omega_{l,t}  \| y_{l,t}^*+\partial f_{l}(y_{l,t}^*)-(W_lx^*_{l-1,t}+b_l)\|^2.
\end{array} \right. 
\end{equation} This completes the proof.

Recall the identity \eqref{r-k-f}, 
$$
r_k=[Id+\partial f_k]^{-1} = [ \partial(\frac{1}{2} \| \,\cdot\,\|^2  + f_k)]^{-1}, \quad k=1,\ldots,40,
$$
and the identity \eqref{r-k-f-1} 
\begin{equation*}
r_k=\partial (\frac{1}{2} \| \,\cdot\,\|^2  + f_k)^*, \quad k=1,\ldots,40,
\end{equation*}
where, as usual, $\psi^*(x)=\sup_{y} [\langle x,y\rangle -\psi(y)] $ is the Legendre-Fenchel conjugate of $\psi$.

\medskip
We have the following lemma whose proof is immediate.
\begin{lemma} \label{mindual} $ $
\begin{itemize}
\item[(i)]\,\,  For any $z$, we have
$$
A_k^{\dagger}[r_k(A_k\theta)-z]= \partial_{\theta} [g_k(A_k\theta)-\langle A_k\theta,  z\rangle],
$$
where $\partial g_k:=(\frac{1}{2} \| \,\cdot\,\|^2  + f_k)^*$.
 \item[(ii)]\,\, \,\,  If there exists $\theta^*$  solution to $r_k(A_k\theta^*)=z$, then it coincides with the minimizer of 
 $\theta \mapsto   g_k(A_k\theta)-\langle A_k\theta,  z\rangle$.
 \end{itemize}
\end{lemma}

\begin{theorem}[Gradient descent algorithm] Suppose the training problem has at least one solution. Then, 
the set of solutions of the training problem coincides with the set of Nash equilibria of the following layer by layer non-zero sum game:
\begin{equation}\label{nash-3}
\arg\min_{\theta_l} \sum_{t=1}^T \omega_{l,t} [g_{l}(A_{l,t}\theta_{l,t})-\langle A_{l,t}\theta_{l,t},  y_{l,t}\rangle],\,\,\, \ l\in \mathcal{L},\,\,\, \omega_{l,t}>0,\ \sum_{t=1}^T \omega_{l,t}=1.
\end{equation}
Moreover, given $\theta^0_{l}:=(\theta^0_{l, 1},\ldots,\theta^0_{l,T})$, the algorithm defined for each $t$ by $\theta^{p+1}_{l,t} =\theta^p_{l,t} - \frac{\gamma}{2 \| A_{l,t}\|^2}A_{l,t}^{\dagger}[r_l(A_{l,t}\theta^p_{l,t})-y_{l,t} ],\,\, 0<\gamma <1$, converges to a minimizer $\theta_l$ of Problem \eqref{nash-3}, as $p\to \infty$.
\end{theorem}

{\bf Proof:}
Since, for each $t=1,\ldots, T$, $ r_l(A_{l,t}\theta)=y_{l,t}$ admits a solution $\theta_{l,t}$ then  by Lemma \ref{mindual} (i) we have 
 $$0\in A_l^{\dagger}[r_l(A_{l,t}\theta_{l,t})-y_{l,t} ] =\partial_{\theta}[g_l(A_{l,t}\theta_{l,t})-\langle A_l\theta_{l,t},  y_{l,t}\rangle], \quad t=1,\ldots, T, 
 $$ 
 with 
 $g_l:=(\frac{1}{2} \| \,\cdot\,\|^2  + f_l)^*$. Therefore, by Lemma \ref{mindual} (ii), $\theta^*_{l,t}$ is a global minimizer of $g_l(A_{l,t}\theta_{l,t})-\langle A_{l,t}\theta_{l,t},  y_{l,t}\rangle$. This in turn yields  that $\theta^*_{l}:=(\theta^*_{l,t})_t$ minimizes the weighted function   
$\sum_{t=1}^T \omega_{l,t} [g_{l,t}(A_{l,t}\theta_{l,t})-\langle A_{l,t}\theta_{l,t},  y_{l,t}\rangle]$.  

Noting that  the inclusion 
$0\in A_{l,t}^{\dagger}[r_l(A_{l,t}\theta^*_{l,t})-y_{l,t} ]$ implies
the inclusion $\theta^*_{l,t} \in \theta^*_{l,t}+ A_{l,t}^{\dagger}[r_l(A_{l,t}\theta^*_{l,t})-y_{l,t} ],$ we examine the fixed-points of the operator  $ \theta \mapsto  \theta+A_{l,t}^{\dagger}[r_l(A_{l,t}\theta)-y_{l,t} ]$. Since $\frac{1}{2 \| A_{l,t}\|^2}(A_{l,t}^{\dagger}\circ r_l \circ A_{l,t})$ is a $\gamma$-averaged operator (see Lemma \ref{lemmann}(ii)), in view of Lemma \ref{lemmann}(vi), for each $t$, given $\theta^0_{l,t}$, the iterates $\theta^{p+1}_{l,t} =\theta^p_{l,t} - \frac{\gamma}{2 \| A_{l,t}\|^2}A_{l,t}^{\dagger}[r_l(A_{l,t}\theta^p_{l,t})-y_{l,t} ], \,0<\gamma <1$, converge to $\theta_{l,t}$, as $p\to\infty$. This completes the proof.

\section{Federated Deep learning}

The Federated learning network has several regional clusters of servers. The set of regional cluster servers is denoted by $\mathcal{S}$ which is a finite (and non-empty) set. Each regional cluster server $s$ interacts with several eligible clients $\mathcal{C}.$ Server $s$ has a neural network model $
\mathcal{M}^s_t:=(R^{s}_{t},W^{s}_t,b^{s}_t,L^{s}, \{\mathcal{H}^{s}_l\}_{1\leq l\leq L^{s}}),
$
that is
$$
R^s_{L^s,t}\circ (W^s_{L^s,t}.+b^s_{L,t})\circ R^s_{L^s-1,t}\circ (W^s_{L^s-1,t}.+b^s_{L^s-1,t})\circ \ldots  \circ R^s_{1,t}(W^s_{1,t}.+b^s_{1,t}),
$$
where $R^s_{l,t}$ is the activation operator of server $s$, $W^s_{l,t}$ is the the weight operator and $b^s_{l,t}$ captures the bias parameter at the server $s$-level. Let  $x^s_0\in \mathcal{H}^s_0,$
$\{\lambda^s_t\}_{t\geq 0}$ be a non-negative sequence and  
$O^s_{l,t}$ given by 
$$
\begin{array}{lll}
O^s_{l,t}:\  & \mathcal{H}^s_{l-1} \rightarrow \mathcal{H}^s_l \\ 
& x^s\mapsto  R^s_{l,t}(W^s_{l,t}x^s+b^s_{l,t})
\end{array}
$$ 
and iterate layer by layer and by timestep the following maps:

\begin{equation} \label{Fed3}
     \begin{array}{l|l}
    \mbox{for}\ t\in \{ 0,1, 2,\ldots\} & 
    y^s_{1,t}=O^s_{1,t}(x^s_t)\\  &
    \mbox{for}\ l\in \{2, \ldots, L\}:\ 
    y^s_{l,t}=O^s_{l,t}(y^s_{l-1,t})\\ &  x^s_{t+1}= x^s_{t}+\lambda^s_t (y^s_{L,t}-x^s_t),
\end{array} 
\end{equation}
which means that  $x^s_{t+1}=x^s_t+\lambda^s_t(O^s_{L,t}\circ O^s_{L-1,t}\circ \ldots \circ O^s_{1,t}(x^s_t)-x^s_t)$ starting from $x^s_0.$ 

The  server  $s$ cannot train such a network on real data $(x_t,y_t)$.  There is a set of eligible clients $\mathcal{C}^s_t\subset \mathcal{C} $ to server $s$ to collaboratively train the network. After selecting and securing the authentication procedure, 
each selected client $c\in \mathcal{C}^s_t$ has $\tau$ time slots to complete a local training on its own device and fully controlled own-data. 
Client $c$ receives  the network architecture model from server $s$ 
$$
(R^{s,c},W^{s,c},b^{s,c},L^{s,c}, \{\mathcal{H}^{s,c}_l\}_{1\leq l\leq L^{s,c}})
$$
and $c$ trains by finding the weights and bias $W^{s,c}_{l,t}, b^{s,c}_{l,t}$ as follows.
Let  $x^{s,c}_0\in \mathcal{H}^{s,c}_0,$
$\{\lambda^{s,c}_t\}_{t\geq 0}$ be a non-negative sequence and  
$O^{s,c}_{l,t}$  given by 
$$ \begin{array}{lll}
O^{s,c}_{l,t}:\ & \mathcal{H}^{s,c}_{l-1} \rightarrow \mathcal{H}^{s,c}_l \\ 
& x^{s,c}\mapsto  R^{s,c}_{l,t}(W^{s,c}_{l,t}x^{s,c}+b^{s,c}_{l,t})
\end{array}
$$ and iterate layer by layer  on $(s,c)$ and by timestep the following:

\begin{equation} \label{Fed4}
     \begin{array}{lll}
\mbox{For}\ t\in \{ 0,1, 2,\ldots\}, \quad 
    y^{s,c}_{1,t}=O^{s,c}_{1,t}(x^{s,c}_t), \\
    \mbox{for}\ l\in \{2, \ldots, L\},\quad 
    y^{s,c}_{l,t}=O^{s,c}_{l,t}(y^{s,c}_{l-1,t}),\\  
    
    \theta^{s,c}_{t+1}= \theta^{s,c}_{t}+\lambda^{s,c}_t ( A^{s,c}_L)^* [r_{L,t}( W_{L,t}^{s,c} y_{L-1,t}^{s,c}+b_{L,t})-y^{s,c}_{L,t}],
\end{array} 
\end{equation}
which means that  $\theta^{s,c}_{t+1}=\theta^{s,c}_t+\lambda^{s,c}_t A_L^* [r_{L,t}( W_{L,t}^{s,c} 
 O^{s,c}_{L-1,t}\circ \ldots \circ O^{s,c}_{1,t}(x^{s,c}_t)-y^{s,c}_{L,t})]$ starting from $\theta^{s,c}_0$, and train the model by updating  the weight and bias $(W^{s,c}_t,b^{s,c}_t)$  to  $(W^{s,c}_{t+1},b^{s,c}_{t+1}).$
The training of the model on $c$ consists of solving the following variational inequality: Given the local input-output (real) data $(x^{s,c}_t,y^{s,c}_t)$ the client finds a vector $\theta^*=(W^{s,c},b^{s,c})$ such that 
$$
\sum_{l=1}^{L^{s,c}} \omega^{s,c}_l  \langle  (r^{s,c}_l[\theta^*]-y^{s,c}_l), A^{s,c}_l(\theta-\theta^*)\rangle \geq 0, \,\,\, \forall x\in \mathcal{H}_0,$$ where $\omega^{s,c}_l>0,\,\,  \sum_{l=1}^l \omega^{s,c}_l=1.
$
The update is 
$\theta_{t+1}= \theta_t +\lambda_t \sum_{l=1}^{L^{s,c}} 
 \omega^{s,c}_l  (A^{s,c}_l)^*(O^{s,c}_l[\theta^*]-y^{s,c}_l),$ with compatible Hilbert spaces.

At  the end of the time slot $t=k,$  the client $c$ sends to the server $s$ 
the trained weight and bias $(W^{s,c}_k,b^{s,c}_k)$ through a secure protocol. The server does not have access to the data  $(x^{s,c}_t,y^{s,c}_t)_t$ of  the client $c$. It is the server $s$ which aggregates the model to $\mathcal{M}^s_{t+1}$
from the collected  partial models $\{(W^{s,c}_k,b^{s,c}_k)\}_{k\leq t, c\in  \mathcal{C}^s_t}$ from eligible clients in $ \cup_{t}\mathcal{C}^s_t.$

\medskip
There are different ways of aggregating the models such as 
\begin{itemize}
\item averaging the parameters: $b_{l,t+1}^s:= \frac{1}{|\mathcal{C}^s_t|} \sum_{c\in \mathcal{C}^s_t}\ b_{l,t}^{s,c},\ $ 
$W_{l,t+1}^{s}:= \frac{1}{|\mathcal{C}^s_t|} \sum_{c\in \mathcal{C}^s_t}\ W_{l,t}^{s,c}, $ for every layer of the network of the server;
 \item averaging the the activation operator $R_{l,t+1}^s:= \frac{1}{|\mathcal{C}^s_t|} \sum_{c\in \mathcal{C}^s_t}\ R_{l,t}^{s,c}$ or a weighted average  $\sum_{c\in \mathcal{C}^s_t}\ \omega^{s,c} R_{l,t}^{s,c}$ with  $\omega^{s,c}\geq 0 ,\ \sum_{c}\omega^{s,c} =1$ 
which is a $\gamma$-averaged operator for some suitable $\gamma\in(0,1]$, by Lemma \ref{lemmann}  (iii).
\end{itemize}

By applying the arguments above, we have

\begin{theorem}
 The outcomes  of the federated neural network $\mathcal{M}^s$  are exactly the  Nash equilibria of the non-zero sum game between regional servers and clients
$\mathcal{G}=(\mathcal{L}^s \cup (\cup_{c\in \mathcal{C}^s} \mathcal{L}^s), 
 (\mathcal{H}_l,  f_l(\cdot)+\frac{1}{2}\| \cdot -b_l-W_lx_{l-1}\|^2)_{l})$
\end{theorem}

\begin{theorem}
The  asymptotic federated training of weights and biases are exactly the solutions of the variational inequality  
involving regional servers and clients. 

\end{theorem}
\section{Illustrative example}
 In the following example we use neural networks to establish the celebrated Gram-Schmidt orthogonalization procedure. 
 We will see that the iteration stops at the second round.

Consider the space of vector-valued random variables $H_0=L^2(\Omega, \mathbb{R}^{d})$ equipped with the topology induced by $\langle x,y\rangle=\mathbb{E}[\text{trace}(x^{\dagger}x)] $ over the probability space $(\Omega, \mathcal{F}, \mathbb{P}) $ and the input data  
$\{ X_0=(x_1,\ldots, x_L)=x\in H_0^L  \}$ of linearly independent random variables $x_k\in H_0,  \ k\in \{1,\ldots, L\}$.

If $x_i$ is non-zero, consider the operator $ P_{x_i}(x_j)$ defined by 
\begin{equation}\label{P-x-i-x-j}
P_{x_i}(x_j):= \left[\frac{ \mathbb{E}[ \text{trace}(x^{\dagger}_j x_i)]}{  \mathbb{E}[ \text{trace}(x^{\dagger}_i x_i)]} \right]x_i.
\end{equation}
The signal $x$ is embedded into $H_1= H_0^L  \times H_0^L $ via the linear map $W_1: x \mapsto (id(x), id(x))= (x,x)$. It is followed by a first activation function 
\begin{equation}\label{P-1}
P_{1}(W_1 x)= ((0, P_{x_1}(x_2),\ldots, P_{x_1}(x_L)) , x)
\end{equation}
followed by a linear map:
$r_1= (a,b) \mapsto (b-a,b)$ to yield $$y_1=r_1[P_{1}(W_1 x)]= ((x_1, x_2-P_{x_1}(x_2),\ldots, x_L-P_{x_1}(x_L), x),$$ which is the output of layer 1.

In layer 2, the weighted operator is 
$$
W_2: H_2=H_1 \rightarrow H_1,\ \ W_2(y_1)=id_{H_1}(y_1)=y_1.
$$
It is followed by an activation operator 
\begin{equation}\label{P-2}
P_2:  \mapsto ((0, 0,P_{y_2}(x_3),P_{y_2}(x_4),\ldots, P_{y_1}(x_L)),x),
\end{equation}
followed by a linear map of the form
$r_2= (a,b) \mapsto (b-a,b)$ to yield  $y_2=r_2[P_{2}(W_2 y_1)]$
which is the output of layer 2.
 
In layer $l$,  
$$
W_l: H_l=H_1 \rightarrow H_1,\ \ W_l(y_{l-1})=id_{H_1}(y_{l-1})=y_{l-1}.
$$
It is followed by an activation operator 
\begin{equation}\label{p-l}
P_l:  \mapsto ((0, 0,\ldots, 0, P_{y_{l}}(x_{l+1}), P_{y_{l}}(x_{l+2}),\ldots, P_{y_{l}}(x_L)),x),
\end{equation}
followed by a linear map of the form
$r_l= (a,b) \mapsto (b-a,b)$ to result to  
$y_l=r_l[P_{l}(W_l y_{l-1})]$ until $L-1$.

At layer $L$, the weight operator is $W_L=H_1 \rightarrow H_1,\ \ W_L(y_{L-1})=id_{H_1}(y_{L-1})=y_{L-1}$
and  $r_L:  H_1 \rightarrow H_0$ is given by 
$$
r_L(y,x)=\left(\frac{y_j}{\sqrt{\mathbb{E} [\text{trace}(y^{\dagger}y)]}}\right )_{j\in \{1,\ldots, L\}}.
$$

In sum, the neural network is given by 
$$
(x_1,\ldots,x_L) \mapsto 
([r_L\circ W_L )] \circ [r_{L-1}\circ P_{L-1}\circ W_{L-1}] \circ [r_{1}\circ P_{1}\circ W_{1}](x)=:O(x).
$$ 

What is the outcome of this network?

\medskip
Let $X_0:=(X_{0}[1],\ldots, X_{0}[L])$ be such that the components $X_{0}[i] \in H_0$ are linearly independent.
In the first round, the neural network provides a random vector $X_1=O(X_0)$. Each component $X_1[i]$ of $X_1$ satisfies $\mathbb{E}[\text{trace}(X^*_1[i]X_1[i])]=1$ which follows directly from the output of the first iteration given by $r_L$. 

We have 
\begin{equation} \label{federatedtraing1u1}
\text{trace}(X^{\dagger}_1[j] X_1[i]) =\delta_{ij}. 
\end{equation}

At the second step of the neural network iteration, we obtain $X_2[1]=X_1[1]$ since $\mathbb{E}[\text{trace}(X^{\dagger}_1[1] X_1[1]]=1$, and 
 $$
 X_2[2]=X_1[2] -\left[\frac{ \mathbb{E}[ \text{trace}(X^{\dagger}_1[2] X_2[1])]}{  \mathbb{E}[ \text{trace}(X^{\dagger}_2[1] X_2[1])]}    \right]    X_2[1]=X_1[2], 
 $$ since $\mathbb{E}[ \text{trace}(X^{\dagger}_1[2] X_1[1])=0$. Hence, the second elements are also identical.
Indeed, by induction, the $j$-th component of $X_2$ reads
$$
X_2[j]= X_1[j] -\sum_{l=1}^{j-1}  \left[\frac{ \mathbb{E}[ \text{trace}(X^{\dagger}_1[j] X_2[l])]}{  \mathbb{E}[ \text{trace}(X^*_2[l] X_2[l])]}    \right]    X_2[l] = X_1[j]. 
$$ Thus, $X_T=X_{T-1}=\ldots =X_1$ for any $T\geq 1.$ We conclude that the output computes a Gram-Schmidt orthonormalization algorithm of any linearly independent family of $L$ random variables.

Let $P_{span\{x\}}$ define the projection onto the  set $span\{x\}$. The map
$$ 
\mathbb{R}\ni m\mapsto \mathbb{E}[\text{trace}(y-m x)^{\dagger}(y-m x)]
$$ is minimized at $m=\frac{\mathbb{E} \text{trace}(x^*y )}{\mathbb{E} \text{trace}(x^{\dagger}x)}$. Thus, $P_{span\{x\}}y= \frac{\mathbb{E} \text{trace}(x^{\dagger}y)}{\mathbb{E}\text{trace}(x^{\dagger}x)}x$ is  the projected random variable $y$ onto the set $span\{x\}$.  Notice that $P_{span\{x\}}$ defines a $1$-Lipschitz i.e. it is $1$-averaged. Furthermore, it coincides with the operator given in \eqref{P-x-i-x-j}.

For $L=2$ layers and the family $X_0=(E[x], x), x\in H_0, \ x\neq 0$ then  
$$X_1=\left( \frac{\mathbb{E}[x]}{\sqrt{\text{trace} (\mathbb{E}[x]^{\dagger}\mathbb{E}[x])}},\  
\frac{x-   \mathbb{E}[x]}{ \sqrt{\mathbb{E}[(x-   \mathbb{E}[x])^{\dagger}(x-   \mathbb{E}[x])]}} \right). $$  
This means that an element of $span(X_1[1], X_1[2])$  is a combination of the  vector
$$
\left( \frac{\mathbb{E}[x]}{\sqrt{\text{trace} (\mathbb{E}[x]^{\dagger}\mathbb{E}[x])}},\  
\frac{x-   \mathbb{E}[x]}{ \sqrt{\mathbb{E}[(x-   \mathbb{E}[x])^{\dagger}(x-   \mathbb{E}[x])]}} \right).
$$
For a given random variable $(y-   \mathbb{E}[y])\in \mathcal{H},$  the projection to $span(X_1[1], X_1[2])$ is 
$$
proj_{span(X_1[1], X_1[2])}(y- \mathbb{E}[y])
= \alpha \frac{\mathbb{E}[x]}{\sqrt{\text{trace} (\mathbb{E}[x]^{\dagger}\mathbb{E}[x])}} 
+\beta  \frac{x-   \mathbb{E}[x]}{ \sqrt{\mathbb{E}[(x-   \mathbb{E}[x])^*(x- \mathbb{E}[x])]}}, 
$$ 
with 
$$\alpha=\langle (y-   \mathbb{E}[y]), X_1[1]\rangle= 0$$
and 
$$\beta= \langle (y-   \mathbb{E}[y]), X_1[2]\rangle=\frac{\mathbb{E}[ (y-   \mathbb{E}[y])^{\dagger}(x-   \mathbb{E}[x])]}{    \sqrt{\mathbb{E}[(x-   \mathbb{E}[x])^{\dagger}(x-   \mathbb{E}[x])]}}=\frac{cov(y,x)}{var(x)},
$$
so that 
$$\mathbb{E}[y]+  proj_{span(X_1[1], X_1[2])} \left( y- \mathbb{E}[y]\right)=\mathbb{E}[y]+\frac{cov(y,x)}{var(x)}(x-   \mathbb{E}[x]).
$$ 
 

 The last expression  $\mathbb{E}[y]+\frac{cov(y,x)}{var(x)}(x-\mathbb{E}[x])$ is linear in $y$ and  is related to the conditional expectation $\mathbb{E}[y | x]$ for the Gaussian random vectors or generally for elliptic random vectors. It differs from conditional expectation in the general setting. By extending the projection activation operator  to the vector case one obtain
$\mathbb{E}[Y]+ \Sigma_{Y,X}\Sigma^{-1}_{X,X} (X-   \mathbb{E}[X])$ where
$\Sigma_{Y,X}= \mathbb{E}[ (Y-   \mathbb{E}[Y])^{\dagger}(X-   \mathbb{E}[X])]$ and $\Sigma_{X,X}=\mathbb{E}[(X-   \mathbb{E}[X])^{\dagger}(X-   \mathbb{E}[X])]$.

\section{The fixed-points of large language models}
Encoder-only transformers, 
Decoder-only transformers, Encoder-Decoder transformers are all compositions of mappings in addition to the above neural network description. Bidirectional Encoder Representations from Transformers, generative pre-trained Transformers use specific architecture for matrix transformations. For example the main decoder architecture is given by   the composition of the attention 
layer  with the feedforward layer iterated in $n$ blocks. $\mathcal{D}= (\sigma_{att_1}\circ \sigma_{ff_1})\circ \ldots  \circ (\sigma_{att_n}\circ \sigma_{ff_n}).$

An attention layer is defined via an an attention multi-head, 
$\sigma_{att} (x)= \mbox{softmax}^*(x \omega_{qk}x') x \omega'_{ov}$  where the autoregressive masking is 
$$\mbox{softmax}^*(A)_{ij}=
 \frac{e^{a_{ij}}}{\sum_{p\geq q} e^{ a_{pq}}} \mathbb{I}_{\{ i\geq j\}},$$
with 
$\omega_{qk}$ is the query-key matrix and $\omega_{ov}$
 the output-value matrix.  By choosing a set of attention head, $AH$ the attention layer is
 $\sigma_{att} (x)=x+\sum_{att\in AH}\sigma_{att} (x).$ The feedforward layer $\sigma_{ff_i}(x)= x+ \sigma_{L}(x).$ The full decoder-only transformer  $$ET(x)=
 \sigma_{e}\circ\mathcal{D} \circ \sigma_u (x),$$ where $\sigma_e$  maps one-hot tokens acts first  via the embedding and  $\sigma_u$  maps the output of the decoder  via the unembedding. All the outputs are normalized via $\sigma_{n,i}(x)=\rho_i+\zeta_i \frac{x_i-\bar{x}}{\sqrt{\nu+\epsilon}}$ where $\epsilon>0, $ $\bar{x}=\frac{1}{size(x)}\sum_{j} x_j$ and $\nu
 =\frac{1}{size(x)-1}\sum_{j} (x_j-\bar{x})^2$, where the $\rho_i$'s and $\zeta_i$'s are parameters. The attention map has the following properties:
 \begin{itemize}
 \item $x\mapsto \sigma_{att} (x)$ is continuously differentiable.
 \item $x\mapsto  \sigma_{ff}\circ  \sigma_{att} (x)$ is $\gamma$-averaged for some $\gamma \in(0,1]$.
 
 \end{itemize}
 
 It follows that the outcomes of large language models are exactly the Nash equilibria of a  game with the new map $  \sigma_n \circ  \sigma_{ff_i}\circ \sigma_{att_i}.$

\section{Conclusion and future works}  \label{sec:4}

In this paper we have studied what is inside the neural network boxes. Both asymptotic network outcome  and training problem were investigated. The methodology is extended to federated neural networks. The algorithms provided here are for $\gamma$-averaged operators on Hilbert spaces. Their convergence is in the weak sense. 

The speed of convergence remains to be investigated. We leave these for future research investigation.

\section*{Acknowledgments}
The work of B. Djehiche is supported by the Verg Foundation. The work of H. Tembine is supported by the TIMADIE MFTG4AI project. The authors would like to thank Jean-Christophe  Pesquet and Patrick Louis  Combettes  for their suggestions on this work.

\end{document}